\documentclass[10pt,english,aps,prl,showpacs,twocolumn]{revtex4}
\usepackage{graphicx}
\usepackage{epsfig}

\begin{document}

\title{Quantum phase transitions of magnetic rotons}
\author{ Joerg Schmalian$^1$ and Misha Turlakov$^2$}
\affiliation{$^{1}$Department of Physics and Astronomy and Ames Laboratory, Iowa State
University, Ames, Iowa 50011, USA \\
$^2$TCM, Cavendish Laboratory, University of Cambridge, Cambridge, CB3 OHE,
UK }
\date{\today}

\begin{abstract}
Due to weak spin-orbit coupling, the magnetic excitations of itinerant
ferromagnet become magnetic rotons, excitations with degenerate minima on a
hypersphere at finite wavevector. Using self-consistent Hartree and
renormalization group calculations, we study \textit{weak fluctuation-driven
first-order quantum} phase transitions, \textit{a quantum tricritical point}
controlled by anisotropy and \textit{non-Fermi liquid behavior} associated
with the large phase volume of magnetic rotons. We propose that magnetic
rotons are essential for the description of the anomalous high-pressure
behavior of the itinerant helical ferromagnet \textit{MnSi}.
\end{abstract}

\maketitle

\vskip2pc The theory of classical, second-order phase transitions has been
extended to quantum phase transitions\cite{Sachdevbook}, especially so in
itinerant electron systems\cite{Hertz,Millis93}. Critical non-Fermi
electron-liquid properties and new phases can emerge near the quantum
critical point, and unconventional behavior of quantum criticality attracts
a lot of current interest\cite{Coleman,Si,Sachdev,Nature,Neutron}. Among
different scenarios for unconventional behavior near the phase transition, a
generic scenario associated with a large phase space of fluctuational modes
has been little explored\cite{Dyugaev}, even though it is well-known for
classical phase transitions\cite{Brazovskii}. A classical phase transition
becomes fluctuation-driven first-order if soft fluctuational modes have a
minimum at a nonzero wave vector $|\mathbf{q}|=q_{0}$ on a hypersphere in $D$%
-dimensional space\cite{Brazovskii}. In the presence of spin-orbit
interaction such soft modes with large phase space appear naturally (see
below) in the proximity of a continuous ferromagnetic transition and can be
called magnetic rotons (in analogy with roton excitations of superfluid $He^{%
\mathrm{4}}$). In this letter we present the theory of \textit{quantum phase
transitions} of magnetic rotons. \textit{\ } 
\begin{figure}[tbp]
\label{modes}\includegraphics[width=7.5cm]{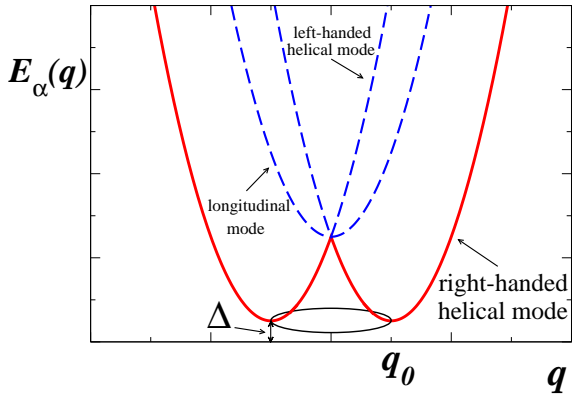}
\caption{The lowest paramagnon mode of a helical magnet becomes, for weak
anisotropy, a magnetic roton with minima for all $\left\vert \mathbf{q}%
\right\vert =q_{0}>0$. The large phase space of magnetic roton fluctuations
is responsible for the unconventional behavior.}
\end{figure}

We show that the large phase volume of magnetic fluctuations with minima at $%
\left\vert \mathbf{q}\right\vert =q_{0}$, independent on the direction of $%
\mathbf{q}$, is responsible not only for the 1$^{\mathrm{st}}$-order nature
of the transition and strong non-Fermi liquid damping of itinerant
electrons, but also for a new tricritical point controlled by anisotropy. We
combine the theory of quantum critical fluctuations, as developed in Ref.%
\cite{Hertz,Millis93}, with the theory of weak \textquotedblleft
crystallization\textquotedblright\ via fluctuation-driven first-order
transitions, as developed in Refs.\cite{Brazovskii,Hohenberg,Dyugaev}. In
addition, the resistivity of itinerant helical ferromagnet and the electron
self-energy are calculated due to the scattering of electrons by roton
fluctuations as well as the strong tendency towards glassy behavior.
Finally, we argue that magnetic rotons form a starting model for the
description of the anomalous high pressure behavior of the itinerant helical
ferromagnet $\mathrm{MnSi}$\cite{Nature}.

We introduce a collective magnetization $\mathbf{M}\left( x,\tau \right) $
along the lines of Refs.\cite{Kataoka,Hertz,Millis93}, valid for a crystal
without inversion symmetry, like $\mathrm{MnSi}$, and obtain the effective
action $S\left[ \mathbf{M}\right] =S_{\mathrm{dyn}}\left[ \mathbf{M}\right]
+S_{\mathrm{stat}}\left[ \mathbf{M}\right] $ which consists of a dynamic
term (see below) as well as the static term 
\begin{eqnarray}
&&S_{\mathrm{stat}}\left[ \mathbf{M}\right] =\int d^{3}\mathbf{x}d\tau
\left( \frac{\delta }{2}\left\vert \mathbf{M}\right\vert ^{2}+\frac{u}{4}%
\left\vert \mathbf{M}\right\vert ^{4}\right)  \nonumber \\
&&+\int d^{3}\mathbf{x}d\tau \left( \frac{\alpha }{2}\sum_{i}\left( \nabla
M_{i}\right) ^{2}+\gamma \mathbf{M}\cdot \left( \nabla \times \mathbf{M}%
\right) \right) .  \label{eq:free-energy}
\end{eqnarray}%
Here, $i=(x,y,z)$ and $\delta \ $determines the proximity to the transition
point. The last term is due to the Dzyaloshinskii-Moriya interaction, caused
by spin-orbit interaction in non-centrosymmetric crystals. The constant $%
\gamma \ll \alpha /a$ is small due to the relativistic origin of \ the
spin-orbit coupling. We assume analyticity of the Landau expansion, Eq.\ref%
{eq:free-energy}, and $\alpha ,u>0$, while other scenarios were suggested%
\cite{Yamada,Belitz,Vojta}. In Eq.\ref{eq:free-energy} we have \ not
included anisotropic terms like $\alpha ^{\prime }\sum_{i}\left( \partial
_{i}M_{i}\right) ^{2}$ and $u^{\prime }\sum_{i}M_{i}^{4}$ allowed by crystal
symmetry, but smaller by $(q_{0}a)^{2}\ll 1$. Quantitative estimates\cite%
{Plumer84} demonstrate that the Dzyaloshinskii-Moriya interaction is indeed
the dominant effect due to spin orbit interaction in \textrm{MnSi.}

Spin-orbit coupling ($\gamma \neq 0$) splits the otherwise degenerate
paramagnons into longitudinal and transverse modes (see the Fig.1). The
energy of the lowest mode, $E\left( q\right) =\Delta _{0}+\left( \left\vert 
\mathbf{q}\right\vert -q_{0}\right) ^{2}/2$, has minima for all $\left\vert 
\mathbf{q}\right\vert =q_{0}$, i.e. is a magnetic roton. The mode
corresponds for $\gamma >0$ to a right handed helical mode. The bare gap $%
\Delta _{0}=\delta -\alpha q_{0}^{2}/2$, which might be tuned by varying
pressure or chemical composition, will be renormalized, $\Delta
_{0}\rightarrow \Delta $, due to interactions. If $\Delta $ is much smaller
than $\frac{\alpha }{2}q_{0}^{2}$, the energy difference between the roton
and the other modes, those modes can be omitted for $k_{\mathrm{B}}T<\frac{%
\alpha }{2}q_{0}^{2}$.

The dynamic part of the action depends on the coupling between magnetic
rotons and particle-hole excitations. If the roton gap $\Delta $ is inside
the particle-hole \textquotedblleft stripe\textquotedblright\ in $(\omega ,%
\mathbf{q})$-space, the damping is linear in frequency\cite{Hertz}: $S_{%
\mathrm{dyn}}=\int_{q}\frac{\left\vert \omega _{n}\right\vert }{\Gamma }%
\left\vert \mathbf{M}\right\vert ^{2}$. Here $\int_{q}=T\sum_{n}\int \frac{%
d^{3}\mathbf{q}}{\left( 2\pi \right) ^{3}}$ and $\Gamma \simeq \frac{%
E_{F}^{2}}{g^{2}}$ with fermion-roton coupling constant, $g$. Since in
general, it is complicated (especially in ordered helical states) to
describe the particle-hole excitations\cite{Moriya} and thus the spectrum
and damping of rotons, we introduce, for simplicity, the variable dynamic
exponent $z$ via $\left\vert \omega \right\vert \rightarrow \ E_{F}\left( 
\frac{\left\vert \omega \right\vert }{E_{F}}\right) ^{2/z}$. $z=2$
corresponds to the over-damped case, while $z=1$ to the undamped case. As we
will show, distinct universality classes occur for different $z$-values,
making it natural to allow $z$ to vary. In fact, $\varepsilon =3-z$ plays a
role similar to the $D-4$ expansion parameter in classical critical
phenomena.

Considering solely fluctuations of the lowest mode we obtain the effective
low energy action 
\begin{equation}
S=\frac{1}{2}\int_{q}\chi _{q}^{-1}\varphi _{q}\varphi
_{-q}+\int_{q_{1}q_{2}q_{3}}\frac{\lambda \left( q_{i}\right) }{4}\varphi
_{q_{1}}\varphi _{q_{2}}\varphi _{q_{3}}\varphi _{q_{4}}  \label{actdiml}
\end{equation}%
with $q_{4}=-q_{1}-q_{2}-q_{3}$ and 
\begin{equation}
\chi _{q}^{-1}\left( r_{0}\right) =|\omega |^{2/z}+r_{0}+(\left\vert \mathbf{%
q}\right\vert -1)^{2}
\end{equation}%
expressed in terms of dimensionless variables: $\mathbf{q}\rightarrow 
\mathbf{q}/q_{0}$ for momenta, $r_{0}=\Delta _{0}/(\alpha q_{0}^{2}/2)$ for
the roton gap, $\omega \rightarrow \frac{\omega }{E_{F}}(2E_{F}/\Gamma
\alpha q_{0}^{2})^{z/2}$ for frequencies and $\lambda \sim uq_{0}^{D+z-4}$
for the coupling constant. Notice that for $D=3$ the coupling constant $%
\lambda \ll 1$ is small for $z>1$, making a weak coupling expansion much
better defined than in the classical limit\cite{Brazovskii}. The momentum
dependence, $\lambda _{\mathbf{q}_{1},-\mathbf{q}_{1},\mathbf{q}_{2},-%
\mathbf{q}_{2}}=\frac{\lambda }{4}\left( 3+\left( \mathbf{q}_{1}\cdot 
\mathbf{q}_{2}\right) ^{2}\right) $ results from the projection onto the
roton mode. It is useful to introduce two coupling constants: $\overline{%
\lambda }$ averaged over the angle between $\mathbf{q}_{1}$ and $\mathbf{q}%
_{2}$ (i.e. for generic angles) and $\lambda _{\parallel }$ where all wave
vectors are parallel\cite{Hohenberg}. \ For a momentum independent initial
interaction, $\lambda _{\parallel }=\frac{8}{7}\overline{\lambda }=\lambda $%
. A weak momentum dependence of $u$ can change $\overline{\lambda }/\lambda
_{\parallel }$ further. Many anomalies related to rotons stem from the fact
that, at low energies, the integrals over momenta are essentially
one-dimensional and that $\lambda _{\parallel }$ renormalizes qualitatively
different from $\overline{\lambda }$.

The nature of the phase transition can be understood from an analysis of the
roton gap $r$ and the interaction vertex $\lambda \left( q_{i}\right) $. A
self-consistent equation for the gap within Hartree approximation, valid for
small $\lambda $, can be obtained from the Dyson equation for the field
propagator: $r=r_{0}+\overline{\lambda }\Sigma _{H}(r),$ with singular
Hartree self-energy,%
\begin{equation}
\Sigma _{H}\left( r\right) =\int_{q}\chi _{q}\left( r\right) =z\frac{\Lambda
^{z-1}-r^{\frac{z-1}{2}}}{2\pi ^{2}\left( z-1\right) },
\end{equation}%
proportional to the local fluctuations of the magnetization, and physical
roton gap, $r$. $\Lambda $ is the upper momentum cutoff. \ For $0<z\leq 1$
the fluctuations of the local magnetization diverge as $r\rightarrow 0$,
like for the classical ($T\neq 0K$) fluctuation-driven 1$^{st}$-order
transition explored by Brazovskii\cite{Brazovskii}. In contrast, $\Sigma
_{H}\left( r\rightarrow 0\right) $ stays finite for $z>1$, relevant for
itinerant magnets, and no disordered state is allowed for $r_{0}<-\overline{%
\lambda }\Sigma _{H}\left( 0\right) $. In the ordered state, the gap of
magnetic rotons 
\begin{equation}
r=r_{0}+\overline{\lambda }\Sigma _{H}(r)+6\overline{\lambda }|m_{0}|^{2},
\end{equation}%
is obtained by adding the Hartree energy due to an helical order parameter
with amplitude $m_{0}$. $m_{0}$ is determined by the equation of state 
\begin{equation}
m_{0}r=3\lambda _{\parallel }\kappa m_{0}{}^{3},
\end{equation}%
with $\kappa =2\overline{\lambda }/\lambda _{\parallel }-1$. For $r_{0}$
smaller than the value $\ r_{0}^{\mathrm{spinodal}}=-\overline{\lambda }%
\Sigma _{H}\left( 0\right) +c_{z}\overline{\lambda }\left( \overline{\lambda 
}\kappa \right) ^{\frac{z-1}{3-z}}$ and $z\leq 3$ three extrema of the
energy exist if $\kappa >0$. $c_{z}$ is of order unity. Decreasing $r_{0}$,
the lowest energy solution jumps from $m_{0}=0$ in the disordered phase to $%
m_{0}\neq 0$ in the ordered phase. The roton mass $r$ discontinuously \
jumps to a larger value in the ordered state. The regime of metastability, $%
r_{0,\mathrm{spinodal}}+\overline{\lambda }\Sigma _{H}\left( 0\right) $,
vanishes as $\kappa \rightarrow 0$. For $\kappa <0$, $m_{0}=0$ is the only
solution of the equation of state, but solutions exist for several ordering
directions of $\mathbf{q}_{0}$ (see below). For $1<z\leq 3$, the phase
transition remains 1$^{\mathrm{st}}$-order for $\kappa >0$, despite the fact
that the Hartree energy does not diverge as $r\rightarrow 0$. It is
important to notice that for $r$ $\rightarrow 0$ the $r$-dependent part of
the self-energy remains the dominant one: $r^{(z-1)/2}\gg r_{0}+\overline{%
\lambda }\Sigma _{H}(0)\ $.

A direct way to demonstrate\ the 1$^{\mathrm{st}}$-order nature of phase
transition is to analyze the renormalized interaction vertex $\ \lambda
_{\parallel }^{r}$. The dominant renormalization of this interaction comes
from \ the polarization \textquotedblleft bubble\textquotedblright\
diagrams. The polarization \textquotedblleft bubble\textquotedblright\ is
defined as $\Pi (r)=\int_{q}\chi _{q}\left( r\right) ^{2}$ and behaves as $%
\Pi (r)=-\frac{\partial \Sigma _{H}\left( r\right) }{\partial r}\sim
r^{(z-3)/2}$. It is necessary to account for a constructive interference
between two channels of zero total momentum\cite{Brazovskii}. Still, the
perturbation series is dominated by terms due to \ $\Pi (r)$ (for a proof
see below). The summation of the leading terms gives: $\lambda _{\parallel
}^{r}=\lambda _{\parallel }\frac{1-\kappa \overline{\lambda }\Pi (r)}{1+%
\overline{\lambda }\Pi (r)}$. If $\Pi (r\rightarrow 0)$\ diverges, $\lambda
_{\parallel }^{r}$ changes sign for $\kappa >0$, implying \ a 1$^{\mathrm{st}%
}$-order transition. \ The renormalized vertex \ $\overline{\lambda }^{r}=%
\frac{\overline{\lambda }}{1+\overline{\lambda }\Pi (r)}$ with generic angle
between momenta behaves differently, and does not change sign. The case $z>3$%
, corresponds to an ordinary 2$^{\mathrm{nd}}$-order phase transition above
the upper critical dimension. In contrast, $z\leq 1$ corresponds to the
dimensionality below the lower critical dimension for second order
transitions. Therefore we have to distinguish between two classes of
fluctuation-driven 1-order transitions characterized by diverging ($z\leq 1$
"classical like") and non-diverging ($1<z\leq 3$ "quantum") local
fluctuations.

In the two-loop approximation, an important difference between an ordinary $%
\phi ^{4}$-theory and the roton field theory becomes evident. Diagrams not
taken into account in the above renormalized vertices, $\lambda _{\parallel
}^{r}$ and $\overline{\lambda }^{r}$, are small by a factor $r^{(D-1)/2}$
which comes from the angle integration if the total momentum of the
propagators inside the \textquotedblleft bubble\textquotedblright\ is not
zero. This implies in case of $\overline{\lambda }=\lambda _{\parallel }$,
where the 1$^{\mathrm{st}}$-order transition occurs if $1-\overline{\lambda }%
\Pi (r)\simeq 0$, or $r^{(3-z)/2}\simeq \lambda $, that self-consistency of
the Hartree solution is guaranteed by the condition $\lambda
^{(D-1)/(3-z)}\ll 1$, much less stringent than for classical systems\cite%
{Hohenberg}.

\begin{figure}[tbp]
\label{FLOW} \includegraphics[width=7.5cm]{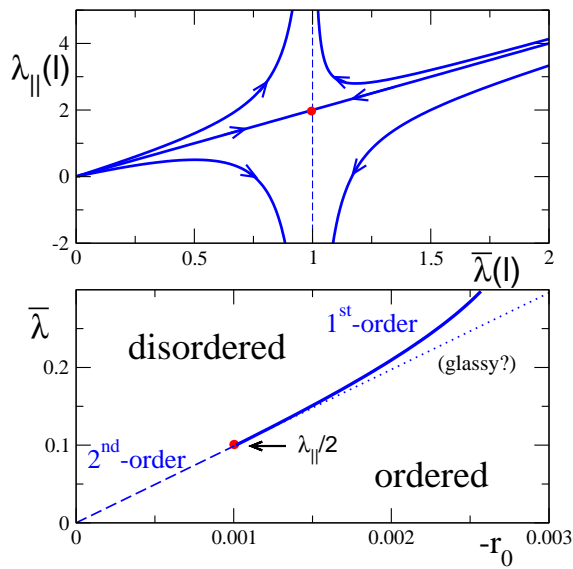}
\caption{{\protect\small (a) Flow diagram for the two coupling constants $%
\protect\lambda_a$ and $\protect\lambda_b$ in units of $\protect\lambda_a^*$%
.(b) Resulting $\protect\lambda_a$-$r_0$ phase diagram for fixed $\protect%
\lambda_b$ with quantum tricritical point at $\protect\lambda_a=2 \protect%
\lambda_b$ separating the regimes with second and first order phase
transition.}}
\end{figure}

In the case $\lambda _{\parallel }=2\overline{\lambda }$ ($\kappa =0$) the
system is right between a regime with a fluctuation induced first order
transition and a regime where the equation of state gives $m_{0}=0$,
suggesting the existence of a new critical end point or a tricritical point.
Insight into the roton model can then be gained by using a renormalization
group approach, following Ref.\cite{Hohenberg,Shankar}. As suggested by our
earlier calculation, the two coupling constants, $\overline{\lambda }$ and $%
\lambda _{\parallel }$, renormalize qualitatively different. We obtain the
flow equations: 
\begin{eqnarray}
\frac{d\overline{\lambda }}{dl} &=&\varepsilon \overline{\lambda }-\overline{%
\lambda }^{2}f\left( Te^{zl},r\left( l\right) \right)   \nonumber \\
\frac{d\lambda _{\parallel }}{dl} &=&\varepsilon \lambda _{\parallel }-2%
\overline{\lambda }^{2}f\left( Te^{zl},r\left( l\right) \right) \ 
\label{flow}
\end{eqnarray}%
together with a corresponding flow equation for $r\left( l\right) $. Here $%
f\left( T,r\right) =3\int_{q}^{>}\chi _{q}^{2}\left( r\right) $ with shell
integration over momenta $\Lambda /e^{l}<\left\vert \mathbf{q-q}%
_{0}\right\vert <\Lambda $. In agreement with the previous paragraph, our
calculation is controlled if $\varepsilon =3-z$ is small. The zero
temperature flow is shown in Fig.2a and is characterized by an unstable
fixed point $\lambda _{\parallel }^{\ast }=2\lambda _{a}^{\ast }\propto
\varepsilon $. The flow away from this fixed point is towards $\lambda
_{\parallel }\rightarrow \pm \infty $ and $\lambda _{a}\rightarrow \lambda
_{a}^{\ast }$. The fixed point can only be reached by tuning $r$ and the
ratio $\frac{\lambda _{\parallel }}{\overline{\lambda }}$, i.e. it is a
tricritical point or \ a critical end point, see also Fig.2b.

The flow towards $\lambda _{\parallel }\rightarrow -\infty $ corresponds to
a fluctuation induced first order transition\cite{Amit}. If $\lambda
_{\parallel }\left( l\right) $ changes sign at the same scale where scaling
stops ($r\left( l\right) =1$) we obtain the bare roton gap at the spinodal, $%
r_{0,\mathrm{spinodal}}$, identical to the result obtained within the Hartee
approach.

Close to the fixed point we can perform a scaling analysis. The free energy
of rotons to first order in $\varepsilon $ is 
\begin{equation}
F\left( \widehat{r},\widehat{\lambda }_{\parallel },T,h\right) =b^{-\left(
1+z\right) }F\left( \widehat{r}b^{1/\nu },\widehat{\lambda }_{\parallel
}b^{\varepsilon },Tb^{z},hb^{y_{h}}\right)
\end{equation}%
where $\nu ^{-1}=2-\frac{\pi }{4}\varepsilon $. $b$ is a scaling parameter
and $\widehat{r}$ and $\widehat{\lambda }_{\parallel }$ are the deviations
of $r$ and $\lambda _{\parallel }$ from their fixed point values,
respectively. $h$ is a field conjugate to the order parameter with $y_{h}=%
\frac{3+z}{2}$. \ Scaling arguments yield for the correlation length $\xi
\left( \widehat{r}\right) \sim \widehat{r}^{-\nu }$, $\xi \left( \widehat{%
\lambda }_{\parallel }\right) \sim \widehat{\lambda }_{\parallel }^{-\frac{1%
}{3-z}}$ and $\xi \left( T\right) \sim T^{-\frac{1}{z}}$, depending on how
one approaches the critical point. \ Varying temperature at the fixed point
gives $C\sim T^{\frac{1}{z}}$ for the singular contribution to the specific
heat and $T_{1}{}^{-1}\sim T^{-\frac{1}{z}}$\ $\ $\ for the NMR spin lattice
relaxation rate. The discontinuity of the order parameter at the first order
transition vanishes like $m\sim \widehat{\lambda }_{\parallel }^{\frac{1}{2}%
\frac{z-1}{3-z}}$as one approaches the fixed point. All these results
require $z>2$ \ such that an additional $w\varphi ^{6}$ interaction is
irrelevant.

Finally, we discuss the flow $\lambda _{\parallel }\left( l\right)
\rightarrow \infty $, which occurs when the bare values obey $\overline{%
\lambda }<2\lambda _{\parallel }$. From the equation of state follows that $%
m_{0}=0$ is the only allowed solution. However, for $z>1$, the renormalized
roton gap vanishes for $r_{0}\rightarrow -\overline{\lambda }\Sigma
_{H}\left( 0\right) $. If $r$ becomes arbitrarily small there will be a
crossover scale where the ignored anisotropy terms, no matter how small,
dominate the low energy physics. States on the hypersphere are not
degenerate anymore, and only isolated points of low energy excitations
become relevant. The universality class is then the quantum version of the
theory analyzed in Ref.\cite{Bak} (see Ref.\cite{Vojta}). For $D=3$ and $z=2$
a mean field second order transition to a state with fixed direction of the
helix occurs.

A crucial difference of the two ordered states reached on either side of the
tricritical point is that the helix direction in the second order case is
determined by anisotropies and therefore fixed, whereas it is arbitrary in
case of the first order transition. In view of \textrm{MnSi}, we propose
that this fixed point is the quantum critical end point\cite{Millis02} of
the observed $\left( p,T\right) =\left( 12\mathrm{kbar},12\mathrm{K}\right) $
tricritical point\cite{Nature} and might be reachable if one varies another
thermodynamic variable like chemical composition.

The equation of state was obtained under the assumption of a single ordering
direction of $\mathbf{q}_{0}$. However, following Ref.\cite%
{Brazovskii,Braz02} one can study more complex choices, like a sum over
helix configuration with varying direction of $\mathbf{q}_{0}$. In case of
large number of distinct directions (i.e. an amorphous superposition of
helix domains) the equation of state changes and we obtain, just like in the
classical case, that these amorphous configurations occur first, at least in
form of metastable solutions. This is consistent with the recent results of
Ref.\cite{SW00} who showed within a replica mean field approach\ to a
similar model that the low temperature state might not be perfectly ordered
but rather be characterized by a distribution of defects of the perfectly
ordered state with glassy properties. Thus in the regime of the $1^{\mathrm{%
st}}$-order transition the system becomes sensitive with respect to the
smallest amounts of random field disorder and forms a self-generated glass.

If the low energy physics is dominated by the quantum tricritical point for $%
\alpha q_{0}^{2}/2\gg kT\gg \Delta =\alpha q_{0}^{2}r/2$, we can also study
the effects on fermions scattered by magnetic rotons. Since the phase volume
of magnetic rotons is large, the non-Fermi-liquid properties are even more
dramatic than close to a 2$^{\mathrm{nd}}$-order quantum phase transitions%
\cite{Hertz,Lonzarich}. Using a local contact coupling with coupling
constant $g$, we find the frequency dependence of the self-energy of fermions%
\cite{Holstein} 
\begin{equation}
\Sigma \left( k_{\mathrm{F}},\omega \right) \simeq \frac{q_{0}g^{2}\Gamma
^{1/2}}{k_{F}^{2}\left( \alpha E_{F}\right) ^{1/2}}\left( \frac{\omega }{%
E_{F}}\right) ^{\frac{z-1}{z}}  \label{eq:sigmafer}
\end{equation}%
(up to a numerical factor), whereas the momentum dependence of the self
energy and the corrections to the fermion-roton vertex are nonsingular. The
absence of singular vertices and the existence of a small coupling constant
are the key reasons why the the non-Fermi liquid nature of the fermions does
not cause feedback onto the collective roton mode, see also Ref.\cite{Abanov}%
. Note, Eq.\ref{eq:sigmafer} was previously obtained in Ref.%
\onlinecite{Dyugaev} for $z=2$. Using the variational approach to the
Boltzmann equation or alternatively a quantum transport theory gives for the
resistivity the result $\rho \propto T^{\frac{z-1}{z}}$, where 
\begin{equation}
\rho \propto \int \int d^{D}kd^{D}k^{\prime }T_{\mathbf{k,k}^{\prime }}\
\left( 1-\cos \theta _{\mathbf{k,k}^{\prime }}\right) 
\end{equation}%
is determined by the angle $\theta _{\mathbf{k,k}^{\prime }}$ between $%
\mathbf{k}$ and $\mathbf{k}^{\prime }$ and the scattering matrix $T_{\mathbf{%
k,k}^{\prime }}$, determined by $Im\chi _{q,\omega }$(see Ref.\cite{Moriya}%
). Away from the quantum tricritical point, this behavior is only expected
at higher temperatures, $kT\gg \Delta $.

Magnetic rotons - magnetic excitations with a minima on a hyper-sphere -
were recently directly observed in a weak itinerant helical ferromagnet $%
\mathrm{MnSi}$ by elastic neutron scattering\cite{Neutron}. In agreement
between theory and experiment the magnetic phase transition is weakly 1$^{%
\mathrm{st}}$-order type in the range of pressures close to critical
pressure and 2-order type away from critical pressure. The resistivity in
the disordered high-temperature phase shows $\sqrt{T}$-behavior\cite{Mena},
which is consistent with our result using $z=2$ for the dynamic exponent due
to conventional Landau damping.

In summary, we have shown that helicoidal magnets with weak anisotropy of
the helix direction display a rich spectrum of interesting properties. We
showed that the system is governed by a quantum tricritical point with
accompanied non-Fermi liquid behavior of the electrons. We propose that the
(dis)ordered state is likely characterized by an amorphous superposition of
domains with different helix directions.

We thank the Aspen Center of Physics for its hospitality. We are grateful
for discussions with A. Chubukov, D. Khmelnitskii, H.v.Lohneysen, G.
Lonzarich, Ch. Pfleiderer, A. Rosch, A. J. Millis and P. G. Wolynes. This
research was supported by an award from Research Corporation, the Ames
Laboratory, operated for the U.S. Department of Energy by Iowa State
University under Contract No. W-7405-Eng-82 (J. S.), and by an EPSRC grant
(M. T.).



\end{document}